
%
%
\font\tenbf=cmbx10

\font\eightrm=cmr8
\font\eightit=cmti8
\font\germ=eufm10
\def\g{\hbox{\germ g}}
\def\sectiontitle#1\par{\vskip0pt plus.1\vsize\penalty-250
 \vskip0pt plus-.1\vsize\bigskip\vskip\parskip
 \message{#1}\leftline{\tenbf#1}\nobreak\vglue 5pt}
\def\wt{\widetilde}
\def\eno{\eqalignno}
\def\ld{\lambda}
\def\ol{\overline}
\magnification=\magstep1
\parindent=15pt
\nopagenumbers
\baselineskip=10pt
\line{
hep-th/930499
\hfil}
\line
{ March 1993 \hfil}
\vglue 5pc
\baselineskip=13pt
\headline{\ifnum\pageno=1\hfil\else%
{\ifodd\pageno\rightheadline \else \leftheadline\fi}\fi}
\def\rightheadline{\hfil\eightit
Generalization of the Gale-Ryser theorem
\quad\eightrm\folio}
\def\leftheadline{\eightrm\folio\quad
\eightit
Anatol N. Kirillov
\hfil}
\voffset=2\baselineskip
\centerline{\tenbf
GENERALIZATION \hskip 0.1cm OF \hskip 0.1cm THE \hskip 0.1cm
GALE - RYSER \hskip 0.1cm THEOREM }
\vglue 24pt
\centerline{\eightrm
ANATOL N. KIRILLOV
}
\baselineskip=12pt
\centerline{\eightit
L.I.T.P., University Paris 7,
}
\baselineskip=10pt
\centerline{\eightit
2 Place Jussieu, 75251 Paris Cedex 05, France
}
\baselineskip=12pt
\centerline{\eightit
and }
\baselineskip=12pt
\centerline{\eightit
Steklov Mathematical Institute,
}
\baselineskip=10pt
\centerline{\eightit
Fontanka 27, St.Petersburg, 191011, Russia
}
\vglue 20pt
\centerline{  }
\centerline{\eightrm ABSTRACT}
{\rightskip=1.5pc
\leftskip=1.5pc
\eightrm\parindent=1pc
We prove an inequality for the Kostka - Foulkes polynomials $K_{\ld ,\mu}(q)$.
As a corollary, we obtain a nontrivial lower bound for the Kostka numbers and
a new proof of the Berenstein - Zelevinsky weight-multiplicity-one-criterium.
\vglue12pt}
\baselineskip=13pt
\overfullrule=0pt
\def\qed{\hfill$\vrule height 2.5mm width 2.5mm depth 0mm$}

\bigbreak

The concept of Young tableau plays an important role in the representation
theory of the symmetric and general linear groups. Based on the pioneering
fundamental works of G. Frobenius, I. Schur, A. Young, H. Weyl and further
developed in the
works of C. Kostka, G. Robinson, A. Richardson, D. Littlewood, C. Schensted,
H. Foulkes, J. Green,
G. James,  M.-P. Schutzenberger, R. Stanley, G .Thomas, A. Lascoux, C. Greene
and many others,
the theory of Young tableaux is now  an important branch of representation
theory and combinatorics with a large number of deep and beautiful
constructions
and results. A good introduction to the subject are the books of D. Littlewood
[L], G. James [J], I. Macdonald [M], B. Sagan [S], W. Fulton [Fu].

An entirely new point of view on the Young tableaux and representation theory
of
general linear and symmetric groups comes from Mathemetical Physics, namely
from the Bethe ansatz [Fa], [FT], [KR]. Bethe ansatz has an important role in
the study
of the exactly solvable models of Mathematical Physics [Fa]. From a
representation theory point of view, the Bethe ansatz (for the $\g l_N$ -
invariant Heisenberg model) gives a very convenient constructive method for
decomposing the tensor
product of irreducible representations (irreps) of the Lie algebra $\g l_N$
into the irreducible parts. In fact, the Bethe vectors appear to be the highest
weight vectors in the corresponding irreducible components. This observation
allows to identify the tensor-product-multiplicities with the number of
solutions of some special system of algebraic equations (Bethe's equations).
Finally, in some particular cases, the number of solutions of the corresponding
Bethe equations admits a combinatorial interpretation in terms of  rigged
configurations [K1], [K3]. On the other hand, it is well-known (see e.g. [L],
[M]), that the multiplicity of an irreducible representation of the Lie
algebra $\g l_N$ in
the tensor product of rectangular-shape-highest-weight irreps may be
identified with the number of Young tableaux of some special kind (e.g.
(semi)standard (super)tableaux, $\ldots$). In this way one can identify a set
of Young
tableaux with a corresponding set of rigged configurations (see e.g. [K1]).

This paper is devoted to the solution of the following problem: given the
partitions $\ld$ and $\mu$, when $\ld$ does only one configuration (see \S 1
below) of the type $(\ld ,\mu )$ exist? This problem may be reformulated in the
following form.
One can prove that for given partitions $\ld $ and $\mu$ there exist an
inequality for the Kostka-Foulkes polynomial $K_{\ld ,\mu }(q)$ (see e.g. [K1],
or \S 2 below):
$$K_{\ld ,\mu }(q)\ge q^c\prod_{n=1}^{\ld_2}\left[\matrix{\sum_{j\le
n}(\mu_j'-\ld_j')+\ld_n'-\ld_{n+1}'\cr
\ld_n'-\ld_{n+1}'}\right]_q,\eqno (0.1)
$$
where $c=n(\ld )+n(\mu )-\sum_n\mu_n'(\ld_n'-1)$.

Here we assume that a $q$-binomial coefficient $\left[ \matrix{m\cr
n}\right]_q$
is equal to zero, if $n\not\in [0,m]$. It is clear that a problem of an
existence of only one configuration of the type $(\ld ,\mu )$ is equivalent to
the following one: to find all partitions $\ld$ and $\mu$ for which the
inequality
(0.1) becomes an equality. The answer is given by the Theorem 2.1. As a
corollary we obtain a simple weight-multiplicity-one-criterium (see Theorem
2.2; compare with [BZ]).

We consider the inequality (0.1) as a generalization of
the Gale-Ryser theorem [R], [M]. Remind that the Gale-Ryser theorem gives a
criterium of an existence of a 0-1 matrix with
given sums of rows and columns:
$$M(e,m)_{\ld'\mu}>0\Longleftrightarrow\ld\ge\mu.\eqno (0.2)
$$
It is well-known (see e.g. [M]) that
$$M(e,m)_{\ld'\mu}=\sum_{\nu}K_{\nu \ld'}K_{\nu'\mu}\ge K_{\ld\mu}
\eqno (0.3)
$$
and
$$\sum_{j\le n}(\mu_j'-\ld_j')\ge 0,~~{\rm for~~all}~~n\ge
1\Longleftrightarrow\ld\ge\mu.
$$
Consequently, from (0.1) - (0.3) we obtain the following nontrivial lower
estimation
$$M(e,m)_{\ld'\mu}\ge\prod_{n=1}^{\ld_2}\pmatrix{\sum_{j\le
n}(\mu_j'-\ld_j')+\ld_n'-\ld_{n+1}'\cr \ld_n'-\ld_{n+1}'}.\eqno (0.4)
$$
It seems to be an interesting problem to construct exactly all Young tableaux
which
correspond to the RHS of inequality (0.1).
\vfill\eject

\bigbreak

{\bf \S 1. Rigged configurations.}
\bigbreak

Let $\lambda$ be a partition and $\mu$ be a composition of some fixed natural
number
$n$. A matrix $m=(m_{k,n})\in M_{l(\lambda )\times l(\mu ')}({\bf Z})$
is called a configuration of type $(\lambda ,\mu)$, if it satisfies
the following conditions
$$\eqalignno{
1)~~~&\sum_{k\ge 1}m_{k,n}=\mu '_n,~~\sum_{n\ge 1}m_{k,n}=\lambda_k,&(1.1)\cr
2)~~~&P_n^{(k)}(m|\mu ):=\sum_{j\le n}(m_{k,j}-m_{k+1,j})\ge 0,&(1.2)\cr
3)~~~&Q_n^{(k)}(m|\mu ):=\sum_{j\ge k+1}(m_{j,n}-m_{j,n+1})\ge 0.&(1.3)}
$$
We denote by $C(\ld ,\mu )$ a set of all configurations of the type
$(\ld ,\mu )$.
Let us define a charge $c(m)$ and cocharge ${\ol c}(m)$ of a configuration
$m$ as follows (see e.g. [LS], [M], [K1]):
$$\eno{
&{\ol c}(m):=\sum_{n\ge 1}\pmatrix{m_{1n}-\mu'_n\cr 2}+\sum_{k\ge 2,n\ge 1}
\pmatrix{m_{kn}\cr 2},\cr
&c(m):=\sum_{k,n}\pmatrix{m_{kn}\cr 2},~~{\rm where}~~\pmatrix{\alpha\cr
2}:={\alpha (\alpha -1)\over 2}.}
$$
At last  for a given configuration $m$ of the type $(\ld ,\mu )$ we define
the following polynomials
$$\eno{
&{\cal K}_m(q)=q^{c(m)}\prod_{k,n}\left[\matrix{P_n^{(k)}(m|\mu
)+Q_n^{(k)}(m|\mu )\cr Q_n^{(k)}(m|\mu )}\right]_q,
\cr
&{\ol {\cal K}}_m(q)=q^{{\ol c}(m)}\prod _{k,n}\left[\matrix{P_n^{(k)}(m|\mu
)+Q_n^{(k)}(m|\mu )\cr Q_n^{(k)}(m|\mu )}\right]_q.}
$$
The following theorem gives an expression for the Kostka-Foulkes polynomial
$K_{\ld ,\mu}(q)$ ($q$-analog of weight multiplicity, see e.g. [LS], [Lu], [M])
as a generating function for rigged configurations.
\medbreak

{\bf Theorem 1.1.} ([K2]).
$$\eno{
&K_{\ld ,\mu}(q)=\sum_{m\in C(\ld ,\mu)}{\cal K}_m(q),&(1.4)\cr
&q^{n(\mu)-n(\ld )}K_{\ld ,\mu}(q^{-1}):={\ol K}_{\ld ,\mu}(q)=\sum_{m\in
C(\ld ,\mu )}{\ol{\cal K}}_m(q).}
$$
It is convenient to imagine a configuration $m\in C(\ld ,\mu )$ as a
collection $\nu$ of partitions (or diagrams)
$\nu =\{ \nu^{(1)},\nu^{(2)},\ldots \}$, where
$$(\nu^{(k)})'_n=\sum_{j\ge k+1}m_{j,n},$$
which satisfy the following conditions
$$\eno{
1)~~&|\nu^{(k)}|=\sum_{j\ge k+1}\ld_j,\cr
2)~~&P_n^{(k)}(\nu |\mu ):=Q_n(\nu^{(k-1)})-2Q_n(\nu^{(k)})+Q_n(\nu^{(k+1)})\ge
0,&(1.6)\cr
&{\rm where}~~\nu^{(0)}:=\mu,~~Q_n(\ld ):=\sum_{j\le n}\ld '_j=\sum_{j\ge
1}\min (n,\ld_j).}
$$
It is clear that
$$\eno{
&P_n^{(k)}(m|\mu )=P_n^{(k)}(\nu |\mu ),\cr
&Q_n^{(k)}(m|\mu )=(\nu^{(k)})_n'-(\nu^{(k)})'_{n+1}.}
$$

{\bf Definition.} Let us call rigged configuration $(\{ \nu \} ;~J)$ a
collection of integer numbers $J:=\{ J_{n,\alpha}^{(k)}\} ,~~1\le\alpha\le
Q_n^{(k)}(m|\mu )$, which satisfy the following conditions
$$0\le J_{n,1}^{(k)}\le J_{n,2}^{(k)}\le \ldots\le J_{n,s}^{(k)}\le
P_n^{(k)}(\nu ~|~\mu ),~~{\rm for~~ all}~~ k,~n.
$$
We assume that the quantum numbers $J_{n,\alpha }^{(k)},~~1\le\alpha\le s:=
Q_n(\nu^{(k)})$ are located in the first column of a set of all length $n$ rows
in
the diagram $\nu^{(k)}$.

Denote by ${\rm QM}(\ld ,\mu )$ the set of all rigged configurations of type
$(\ld ,\mu )$.
\medbreak

{\bf Theorem 1.2.} ([K1]). There exist a natural bijection between the set
STY($\ld ,
\mu $)of all (semi) standard Young tableaux of a shape $\ld$ and weight
$\mu$ and QM($\ld ,\mu$):
$${\rm STY}(\ld ,\mu )\rightleftharpoons {\rm QM}(\ld ,\mu ).
$$

{\bf Corollary 1.3.} ({\it Maximal configuration}). Let us assume that
$\ld\ge\mu$ with respect to the dominant order (see e.g. [M]).
Consider the matrix $m=(m_{k,n})$, where
$m_{k,n}:=(\mu'_n-\ld'_n)\delta_{k,1}+\theta (\ld'_n-k)$,
and
$$\theta (x)=\left\{\matrix{1,~~{\rm if}~~x\ge 0,\cr 0,~~{\rm if}~~x<0.}
\right.
$$
Then $m\in C(\ld ,\mu )$.

The proof is an easy consequence of the following
inequalities
$$\eno{
&P_n^{(k)}(m|\mu )=[Q_n(\mu )-Q_n(\ld )]\delta_{k,1}+\min
(\ld_k,n)-\min(\ld_{k+1},n)\ge 0,\cr
&Q_n^{(k)}(m|\mu )=\max (\ld'_n,k)-\max (\ld'_{n+1},k)\ge 0.}
$$
\qed

It is clear that the configuration under
consideration corresponds to the following collection of diagrams
$$\{ \ld [1],\ld [2],\ldots \} ,
$$
where the partitions $\lambda[k],~~k\ge 1$, are defined as follows
$$
(\ld [k])_n=\ld_{k+n},~~n\ge 1.
$$
We will call this configuration  {\it the maximal configuration} of type $(\ld
,\mu )$ and denote
it by $\Delta$.
\medbreak

{\bf Corollary 1.4.} If $ \nu \in C(\ld ,\mu )$, then
$$P_n^{(1)}(\nu |\mu )\le Q_n(\mu )-Q_n(\ld ).
$$

Proof. Let us assume the converse, namely, that there exist $n\ge 1$ such that
$$\eno{
&P_n^{(1)}(\nu |\mu )>Q_n(\mu )-Q_n(\ld ),~~{\rm or~~equivalently,}\cr
&P_n^{(1)}(\nu |\ld )=Q_n(\ld )-2Q_n(\nu^{(1)})+Q_n(\nu^{(2)})>0.}
$$
Let us note that $P_n^{(k)}(\nu |\ld )\ge 0$ for all $k\ge 2$ and $n\ge 1$.
So the set
${\rm QM}(\ld ,\ld )$ contains at least two elements $\Delta$ and $\nu$, which
contradicts the well-known fact $|{\rm STY}
(\ld ,\ld )|=1$.

\qed
\medbreak

{\bf Corollary 1.5.} $C(\ld ,\mu )\ne \phi\Longleftrightarrow \ld \ge\mu$.

Proof. If $\ld\ge\mu$, then $\Delta\in C(\ld ,\mu )$. Now let us consider a
configuration
$\nu\in C(\ld ,\mu )$. If $\ld$ does not dominate $\mu$, then we have
$Q_n(\mu )-Q_n(\ld )<0$ for some $n$,  and consequently (see Corollary 1.4),
$P_n^{(1)}(\nu |\mu )<0$, which
is a contradiction with condition (1.2).

\qed
\medbreak

{\bf Corollary 1.6.} If $\nu\in C(\ld ,\mu )$, then $\ld [k]\ge \nu^{(k)}$.
In particular, $\nu_1^{(k)}\le\ld_{k+1}$ for
all $k\ge 1$.

Proof. Let us consider the diagram $\ld [k]$ and a collection of partitions
${\wt \nu}=\{ \nu^{(k+1)},\nu^{(k+2)},\ldots \}$. It is clear, that
${\wt\nu}\in C(\ld [k],\nu^{(k)})$. Consequently, $\ld [k]\ge\nu^{(k)}$.
\qed
\medbreak

{\bf Corollary 1.7.} If $\nu\in C(\ld ,\mu )$, then
$$P_n^{(k)}(\nu |\mu )\ge \min (\ld_k,n)-\min (\ld_{k+1},n).
$$
\medbreak

Note, that we may rewrite a definition of ${\ol c}(\nu )$ in the  following
form
$${\ol c}(\nu )=\sum_{n\ge 1}\pmatrix{-\alpha_n^{(1)}\cr 2}+\sum_{k\ge 2,n\ge
1}\pmatrix{\alpha_n^{(k-1)}-\alpha_n^{(k)}\cr 2}-n(\ld ),
$$
i.e. the cocharge ${\ol c}(\nu )$ of a configuration $\nu\in C(\ld ,\mu )$
depends
only on the
configuration $ \nu $ and does not depends on the composition $\mu$. Here
$\alpha_n^{(k)}:=(\nu^{(k)})_n'$.
\bigbreak
\vskip 0.5cm

{\bf \S 2. Generalization of the Gale-Ryser theorem.}
\bigbreak

{}From an existence of the maximal configuration $\Delta$ of the type
$(\ld ,\mu )$ it follows that
$$K_{\ld ,\mu }(q)\ge{\cal K}_{\Delta}(q)=q^{c(\Delta )}\prod_{n=1}^{\ld_2}
\left[\matrix{Q_n(\mu )-Q_n(\ld )+\ld_n'-\ld_{n+1}'\cr
\ld_n'-\ld_{n+1}'}\right]_q,\eqno (2.1)
$$
where
$$c(\Delta )=\sum_{n\ge 1}\pmatrix{\mu_n'-\ld_n'\cr 2}=n(\ld )+n(\mu )-\sum_n
\mu_n'(\ld_n'-1).
$$
Note, that ${\rm deg}~{\cal K}_{\Delta}(q)=n(\mu )-n(\ld )$, where $n(\ld
):=\sum_{i\ge 1}(i-1)\ld_i$.

We will study the question for which partitions $\ld$ and $\mu$ the  unequality
(2.1)
becomes an equality. This
exactly means that there exist only one configuration.
\medbreak

{\bf Theorem 2.1.} There exist only one configuration of the type
$(\ld ,\mu ),~~\ld \ge \mu$, if and only if the following conditions are valid:

i) $\ld_2=1$, i.e. $\ld$ is a hook,

ii) if $\ld_2\ge 2$, then
for all $1\le n_0<n_1\le\ld_2,~~(\ld_0:=+\infty )$, such that
$\ld_{n_0-1}'>\ld_{n_0}'\ge \ld_{n_1}'>\max
(\ld_{n_1+1}',~1)$, we have either
$$\eno{
&Q_{n_0}(\mu )-Q_{n_0}(\ld )\le 1,~~{\rm or}& (2.2)\cr
&Q_{n_1-1}(\mu )-Q_{n_1-1}(\ld )\le 1.}
$$

Proof. At first, let us prove the necessity of condition (2.2).
Let us consider a
perturbated configuration
$${\wt
m}_{kn}=m_{kn}-a(\delta_{kk_0}-\delta_{kk_1})(\delta_{nn_0}-\delta_{nn_1}),
$$
where $1\le n_0<n_1\le\ld_2,~~1\le k_0<k_1$. From a simple calculation it
follows that
$$\eno{
&{\wt P}_n^{(k)}({\wt m})=P_n^{(k)}(m)-a(\delta_{kk_0}-\delta_{k+1,k_0}-
\delta_{kk_1}+\delta_{k+1,k_1})\chi (n\in [n_0,n_1)),\cr
&{\wt Q}_n^{(k)}({\wt m})=Q_n^{(k)}(m)+a(\delta_{nn_0}-\delta_{n+1,n_0}-
\delta_{nn_1}+\delta_{n+1,n_1})\chi (k\in [k_0,k_1)),& (2.3)\cr
&c({\wt m})=c(m)-a(m_{k_0n_0}-m_{k_0n_1}-m_{k_1n_0}+m_{k_1n_1}-2a).}
$$
Now let us take $m$ to be the maximal configuration 
of the type
$(\ld ,\mu )$. It is clear from (2.3) that a pertubartion ${\wt m}$ of a
maximal
configuration would exist only if $k_0=1$ and $k_1=2$, and then
$$\eno{
&{\wt P}_n^{(1)}=P_n^{(1)}(\Delta )-2\chi (n\in [n_0,n_1)),\cr
&{\wt P}_n^{(2)}=P_n^{(2)}(\Delta )+\chi (n\in [n_0,n_1)),&(2.4)\cr
&({\wt \nu}^{(1)})_n=\max (\ld_n'-1,0)+\delta_{nn_0}-\delta_{nn_1}.}
$$
Here we use Garsia's notation
$$\eqalignno{
&\chi (P)=1,~~{\rm if}~~P~~{\rm is~~true},\cr
&\chi (P)=0~~{\rm otherwise}}
$$
So, if the condition (2.2) is not valid then there exist indices $n_0,
{}~~n_1,~~ 1\le n_0<n_1\le\ld_2$ such that
$$\eno{
i)~~&\ld_{n_0-1}'>\ld_{n_0}\ge\ld_m'>\max (\ld_{n_1+1}',1)\cr
ii)~~&Q_n(\mu )-Q_n(\ld )\ge 2,~~{\rm for~~all}~~n\in [n_0,n_1).}
$$
It follows from (2.4) that a pertubarated configuration
$${\wt\Delta}_n^{(k)}=\Delta_n^{(k)}-(\delta_{nn_0}-\delta_{nn_1})(\delta_{k1}
-\delta_{k2})
$$
belongs to the set $C(\ld ,\mu )$.

Secondly, let us check the sufficiency of condition (2.2); thus it is
needed to prove that
under condition (2.2) there exist only one configuration of type $(\ld ,\mu )$.
For this goal let us use the following inequalities (see (1.6)):
$$P_n^{(r)}(\nu |\mu )=Q_n(\nu^{(r-1)})-2Q_n(\nu^{(r)})+Q_n(\nu^{(r+1)})\ge 0.
$$
Multiplying this inequalities on $r$ and summing up till some fixed $k$, we
obtain an inequality
$$Q_n(\mu )-(k+1)Q_n(\nu^{(k)})+kQ_n(\nu^{(k+1)})\ge 0.\eqno (2.5)
$$
Now let us take $l=l(\ld )\ge 2$, and $p=\ld_l$. Then we have
$\nu^{(l)}=\phi$ and
$$\eno{
&Q_n(\mu )-lQ_n(\nu^{(l-1)})\ge 0,~~{\rm or~~equivalently,}\cr
&Q_n(\ld )-lQ_n(\nu^{(l-1)})\ge -[Q_n(\mu )-Q_n(\ld )].&(2.6)}
$$
But if $n\le p$, then it is clear that
$$
Q_n(\ld )=l\min (n,p)=lQ_n(\Delta^{(l-1)})
$$
and consequently, we may rewrite (2.6) as follows:
$$Q_n(\Delta^{(l-1)})-Q_n(\nu^{(l-1)})\ge -{Q_n(\mu )-Q_n(\ld )\over l}.\eqno
(2.7)
$$
Now let us show using (2.2) and (2.7), that $\nu^{(l-1)}=\Delta^{(l-1)}$.
This is evident if
$p:=\ld_l=1$. If we have $p\ge 2$, then the condition (2.2) with $n_0=1$ and
$n_1=p$  means that either
$Q_1(\mu )-Q_1(\ld )\le 1$, or $Q_{p-1}(\mu )-Q_{p-1}(\ld )\le 1$. In the
first case we have $1-(\nu^{(l-1)})'_1\ge -\displaystyle{{1\over l}}$, or
equivalently,
$(\nu^{(l-1)})'_1=1$
and, consequently, $\nu^{(l-1)}=(1^p)=\Delta^{(l-1)}$. In the second case, if
we assume $\nu^{(l-1)}\ne \Delta^{(l-1)}$, then
$\nu_1^{(l-1)}<p$, and hence (using (2.7)) $\min (p-1,p)-p\ge -
\displaystyle{{1\over l}}$, but
this is impossible. Thus we proved that $\nu^{(l-1)}=\Delta^{(l-1)}$.

Now we use an induction. So, let us assume that $\Delta^{(r)}=\nu^{(r)}$, when
$k+1\le r\le l-1$. We must prove that $\Delta^{(k)}=\nu^{(k)}$.
Note that it follows from an equality $\Delta^{(k+1)}=\nu^{(k+1)}$ that if
$n\le\ld_{k+1}$ then:
$$\eno{
&Q_n(\nu^{(k+1)})=Q_n(\Delta^{(k+1)})=Q_n(\ld )-n(k+1),\cr
&Q_n(\Delta^{(k)})=Q_n(\ld )-kn.}
$$
Consequently, using (2.5) we find
$$\eno{
0&\le Q_n(\mu )-(k+1)Q_n(\nu^{(k)})+kQ_n(\nu^{(k+1)})=\cr
&=Q_n(\mu )-Q_n(\ld )+Q_n(\ld )-(k+1)Q_n(\nu^{(k)})+kQ_n(\ld )-k(k+1)n=\cr
&=Q_n(\mu )-Q_n(\ld )+(k+1)[Q_n(\ld )-kn-Q_n(\nu^{(k)})]=\cr
&=Q_n(\mu )-Q_n(\ld )+(k+1)[Q_n(\Delta^{(k)})-Q_n(\nu^{(k)})].}
$$
Hence we have
$$Q_n(\Delta^{(k)})-Q_n(\nu^{(k)})\ge-{Q_n(\mu )-Q_n(\ld )\over k+1},~~1\le n
\le \ld_{k+1}.\eqno (2.8)
$$
Note, that from Corollary 1.6 it follows an inequality
$$Q_n(\Delta^{(k)})-Q_n(\nu^{(k)})\le 0.\eqno (2.9)
$$
Further, using Corollary 1.6 
and an induction assumption, one
can easily prove that
$$(\nu^{(k)})'_n=(\Delta^{(k)})'_n,~~{\rm if}~~1\le n\le\ld_{k+2}.
$$
Now let us use inequality (2.8) and condition (2.2) when $\ld_{k+2}\le n\le
\ld_{k+1}$, where $\ld_{k+2}=n_0+1$ and $\ld_{k+1}=n_1$. We may assume, that
$\ld_{k+1}-\ld_{k+2}\ge 2$. If we have $Q_{n_1-1}(\mu )-Q_{n_1-1}(\ld )\le 1$
and $\nu_1^{(k)}<\ld_{k+1}=n_1$, then from (2.8) it follows
$$Q_{n_1-1}(\Delta^{(k)})-Q_{n_1-1}(\nu^{(k)})=-(\nu^{(k)})'_{n_1}\ge 0,$$
but according to (2.9)
this is possible only if $Q_{n_1-1}(\Delta^{(k)})=Q_{n_1-1}(\nu^{(k)})$, and
consequently, $\Delta^{(k)}=\nu^{(k)}$.
By the same reasons, if $Q_{\ld_{k+2}+1}(\mu )-Q_{\ld_{k+2}+1}(\ld )\le 1$
and $\epsilon :=(\nu^{(k)})'_{\ld_{k+2}+1}\ge 2$, then we  have (using (2.8)):
$$Q_{n_0}(\Delta^{(k+1)})+\ld_{k+2}+1-Q_{n_0}(\Delta^{(k+1)})-\ld_{k+2}-
\epsilon\ge 0,$$
and, consequently, $\epsilon \le 1$. This is a contradiction with our
assumption that $\epsilon\ge 2$. Consequently, $\epsilon =1$ and
$\Delta^{(k)}=\nu^{(k)}$.

\qed

Now let us consider a weight-multiplicity-one problem (see [BZ]). An answer
has been obtained by A.Berenstein and A.Zelevinsky [BZ]. We assume to give a
weight-multiplicity-one-criterium as corollary of Theorem 2.1. One can easily
show that our criterium is equivalent to the Berenstein-Zelevinsky one. Thus we
want to answer the question: when is the Kostka number $K_{\ld ,\mu}$ equal
to 1? We may assume that $\mu$ is a partition and $\ld_1'=\ldots =\ld_{n_1}'>
\ld_{n_1+1}'=\ldots =\ld_{n_2}'>\ld_{n_2+1}'=\ldots >\ld_{n_{k-1}+1}'=\ldots
=\ld_{n_k}'>0$.

Let us denote by $\ld^{(l)}$ and $\mu^{(l)},~~1\le l\le k$, the following
partitions
$$\eqalignno{
&\ld^{(l)}:=(\ld_{n_{l-1}+1}',\ldots ,\ld_{n_l}')~~{\rm
of~~rectangular~~shape,}\cr
&\mu^{(l)}:=(\mu_{n_{l-1}+1}'\ldots ,\mu_{n_l}'),}
$$
where we assume $n_0:=0$.
\medbreak

{\bf Proposition 2.2.} (weight-multiplicity-one-criterium).

The Kostka number $K_{\ld ,\mu},~~\ld\ge\mu$, is equal to 1 if, and only if,
the
following conditions are valid

i) $ \ld^{(l)}\ge\mu^{(l)}$,
{\hfill$(2.10)$}

\hskip-0.5cmwith respect to the dominant order on partitions (see e.g. [M]); in
partiqular $|\ld^{(l)}|=|\mu^{(l)}|$,\break $1\le l\le k$;

ii) for all $1\le l\le k$ we have either
$$\eqalignno{
&0\le\mu_{n_{l-1}+1}'-\ld_{n_{l-1}+1}'\le 1,~~{\rm or}\cr
&0\le\ld_{n_l}'-\mu_{n_l}'\le 1.& (2.11)}
$$
Proof. It is clear that $K_{\ld ,\mu}=1$ iff there exist only one configuration
$\Delta$ (a maximal one) and ${\cal K}_{\Delta}(1)=1$ (see Theorem 1.1). The
condition (2.10) is equivalent to ${\cal K}_{\Delta}(1)=1$. The condition
(2.11) follows from Theorem 2.1.

\qed

If a partition $\ld :=(n^m)$ has a rectangular shape, then we have two typical
examples for $K_{\ld ,\mu}=1$. Namely,
$$\eqalignno{
i)~~&\ld =(n^m),~~\mu\subset (n^{m+1}),~~\ld\ge\mu;& (2.12)\cr
ii)~~&\ld =(n^m),~~\mu=(n^{m-1})\cup{\wt\mu},~~{\wt\mu}\vdash n,~~\ld\ge\mu.}
$$
According to Proposition 2.2, a general example with $K_{\ld ,\mu}=1$ may be
glued from the elementary examples (2.12).

{\bf Examples.}
$$\eqalignno{
&K_{\ld ,\mu}(q)=q^7,~~{\rm if}~~\ld =(7,7,7),~~\mu =(6,6,5,4),\cr
&K_{\ld ,\mu}(q)=q^{6n},~~{\rm if}~~\ld =(4n,4n,4n),~~\mu =(4^{3n}),~~n\ge 1.}
$$
\bigbreak
\vskip 0.5cm

{\bf Acknowledgements.} I am gratefully acknowledge to L.D. Faddeev,
A. Lascoux, A. Berenstein and J.-Y. Thibon for helpful discussions and kind
support during the preparation of this paper.
\vskip 1cm

{\bf References.}
\vskip 0.5cm

\item{[BZ]} Berenstein A.D., Zelevinsky A.V. When is the multiplicity of a
weight equal to 1? Funct. Anal. i ego Prilozheniya, 24, 1990, n. 4, 1-13.
\item{[Fa]} Faddeev L.D., Integrable models in $1+1$ dimensional quantum field
theory, in Les Houches Lectures 1982, Elsevier, Amsterdam, 1984.
\item{[FT]} Faddeev L.D., Takhtajan L.A. Spectrum and scatering of excitations
in one-dimensional isotropic Heisenberg magnet. Zap. Nauch. Semin. LOMI, 109,
1981, 134-173 (in Russian).
\item{[Fu]} Fulton W. Young Tableaux (with applications to representation
theory and geometry). Chicago, 1992.
\item{[J]} James G. The representation theory of the symmetric groups. Lecture
Notes in Math., 682, Springer-Verlag, 1978.
\item{[K1]} Kirillov A.N. On the Kostka-Green-Foulkes polynomials and
Clebsch-Gordan numbers. Journ. Geom. and Phys., 5, 1988, 365-389.
\item{[K2]} Kirillov A.N. Decomposition of symmetric and exterior powers of the
adjoint representation of $gl_N$. RIMS preprint, 840, 1991.
\item{[K3]} Kirillov A.N. Completeness of the Bethe vectors for generalized
Heisenberg magnet. Zap. Nauch. Semin. LOMI, 134, 1984, 169-189 (in Russian).
\item{[KR]} Kirillov A.N., Reshetikhin N.Yu. Yangians, Bethe-ansatz and
combinatorics. Lett. in Math. Phys., 12, n. 7, 1986, 199-208.
\item{[LS]} Lascoux A., Schutzenberger M.-P. Sur une conjecture de H.O.Foulkes.
C.R. Acad. Sci. Paris, 286A, 1978, 323-324.
\item{[L]} Littlewood D. The theory of group characters. 2nd edn. Oxford
University Press, 1950.
\item{[Lu]} Lusztig G. Singularities, character formulas, weight
multiplicities. Asterisque, 101-102, 1983, 208-229.
\item{[M]} Macdonald I. Symmetric functions and Hall polynomials. Oxford
Mathematical Monographs, 1979.
\item{[R]} Ryser H. Combinatorial mathematics. Wiley, 1963.
\item{[S]} Sagan B. The Symmetric group. Wadsworth, Belmont, CA, 1991.

\end